# ENHANCED COUPLING DESIGN OF A DETUNED DAMPED STRUCTURE FOR CLIC

A. D'Elia[†*‡], R.M. Jones[†*], A. Grudiev[‡], T. Higo[°], V.F. Khan[‡], W. Wuensch[‡]
†School of Physics and Astronomy, The University of Manchester, Manchester, U.K.
*The Cockcroft Institute of Accelerator Science and Technology, Daresbury, U.K.
‡CERN, Geneva, Switzerland.
°KEK, Tsukuba, Japan.

*Abstract*

The key feature of the improved coupling design in the Damped Detuned Structure (DDS) is focused on the four manifolds. Rectangular geometry slots and rectangular manifolds are used. This results in a significantly stronger coupling to the manifolds compared to the previous design. We describe the new design together with its wakefield damping properties.

## INTRODUCTION

The main issue entailed in the design of high gradient structures for CLIC [1], [2], [3] is to provide efficient transfer of rf energy to the accelerating beams. In addition, beam excited long range wakefields must be properly suppressed whilst ensuring the fields on the surfaces of the cavity are sufficiently low to prevent breakdown. These two requirements are not independent and are usually antithetical.

The present CLIC baseline design relies on heavy damping ($Q\sim10$ [4]) which is achieved through strong coupling of the beam excited wakefield via waveguides attached to each cell. This damping scheme is efficient and also recent high power tests are encouraging [5]. Nonetheless to efficiently dissipate the power radiated in the loads, the radial size of the disks cannot be made too small and this may have an impact on the overall cost of the structure. In parallel to this *waveguide damping* scheme other alternatives are under investigations such as choke mode damping [4] and the DDS scheme outlined here.

The latter is based on the experience of NLC [7], [8] and entails detuning the cell frequencies of the first dipole band in an error function fashion, by tapering down the irises along the structure. This frequency spread prevents the wake from adding in phase. However, due to the limited number of cells in the structure the wakefield eventually recoheres. The recoherence can be suppressed by providing a moderate coupling ($Q\sim500$-$1000$ for the NLC) to four attached manifolds running parallel to the beam axis. Interleaving neighbouring structure frequencies enhances wakefield suppression.

A prototype structure, CLIC_DDS_A, which will be submitted to high power tests (71MW peak input power) has been studied [9], [10], [11], [12], designed [13] and is presently under fabrication in Japan, under the supervision of KEK. The fabrication of this structure is expected to be finished by the second quarter of 2012.

The coupling between cell and manifolds is achieved with a key-hole shaped slot, cut on the iris wall. This slot perturbs the surface H-field pattern and is a limitation in the maximum achievable coupling ($Q\sim1700$). This in turn limits the wakefield damping also.

In the following we will discuss a new cell design which enhances the coupling between cells and manifolds and also simplifies the mechanical design. A comparison between the wakefield of CLIC_DDS_A and of the new design will be shown. Finally we will briefly discuss the potential for this structure to be used for the CLIC main linac.

## SINGLE CELL DESIGN

Figure 1 shows the H-field computed with HFSS (1/8$^{th}$ of the full geometry) of previous DDS (left side) in comparison with the new design. In the earlier DDS design a circular manifold is attached to a key-hole slot which penetrates deeply inside the cell: this makes the geometrical design more intricate and in addition a significant perturbation of the surface fields occurs. In the latest design a rectangular manifold is attached via a rectangular slot to the cell. This design reduces the surface H-field by ~20% compared to the previous design (see table in Fig. 1). The R/Q is essentially unmodified (20.38kΩ/m compared to 20.46kΩ/m).

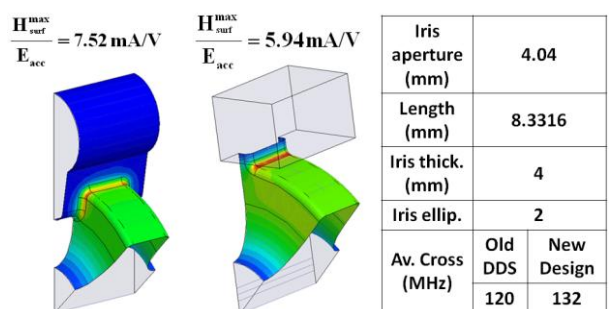

Figure 1: Surface H-Field distribution in earlier DDS (left) and surface H-Field distribution in the latest design (right). The table illustrates the geometrical parameters used in the simulation and avoided crossing values for the two designs.

It is worth noticing that the earlier DDS exhibited an H-Field enhancement in the vicinity of the key-hole slot and this is difficult to alleviate. In practice this limits the achievable coupling. The field enhancement on the outer wall of the cell can be instead reduced by making the outer wall elliptical, as was done in CLIC_DDS_A cells ($H_{max}/E_{acc}$~4.9mA/V, corresponding to a pulsed temperature rise of ~51 K).

In the further studies we explored the wakefield damping capability of the new design. The outer wall of the cell stays circular and we chose an iris aperture of 4.04mm for the first cell and 1.94mm for the last one. We have first focused on the uncoupled cell, i.e. without manifold, to choose optimum thickness and ellipticity of the irises in terms of group velocity ($v_g$) and bandwidth ($\Delta\omega/2\pi$), i.e. maximum difference of the synchronous frequencies ($\omega_{syn}/2\pi$) of these two extreme cells.

The results are shown in Fig. 2.

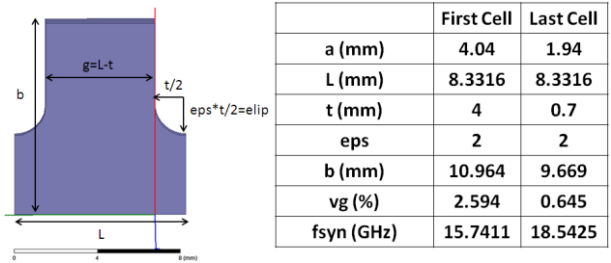

|  | First Cell | Last Cell |
|---|---|---|
| a (mm) | 4.04 | 1.94 |
| L (mm) | 8.3316 | 8.3316 |
| t (mm) | 4 | 0.7 |
| eps | 2 | 2 |
| b (mm) | 10.964 | 9.669 |
| vg (%) | 2.594 | 0.645 |
| fsyn (GHz) | 15.7411 | 18.5425 |

Figure 2: Geometrical parameters of the first and last cell.

The second step has been to analyze the dependence of the coupling (in this particular case we analyzed the avoided crossing) on the cell geometrical parameters when coupled with the manifolds. The results are illustrated in Fig. 3: $W_W$ and $W_H$ do not significantly modify the figures of merit and we have maintained them at 5mm and 6mm, respectively. The coupling is mainly determined by the width ($W_S$) and height ($H_S$) of the slot. For CLIC_DDS_A we had, after optimization, an avoided crossing of ~200MHz corresponding to a Q~1650. In all the cases reported in Fig. 3, the R/Q of the monopole stayed practically unchanged ranging from 19.4 to 20.4k$\Omega$/m.

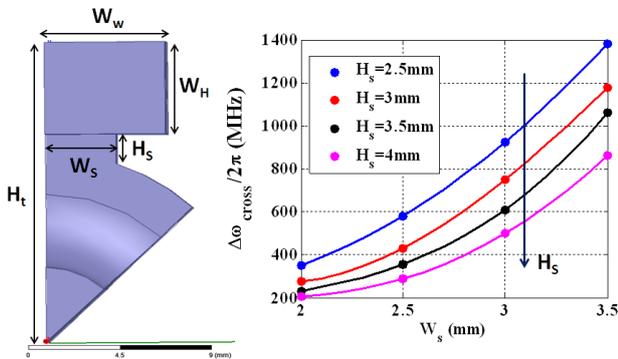

Figure 3: Dependence of the average avoided crossing separation, $\Delta\omega_{cross}/2\pi$, on the geometrical parameters.

The other important parameter is $\Delta\omega/2\pi$ (see Fig. 4). We have decided to keep $H_t$ constant throughout the structure to simplify the mechanical design. Also $W_S$ has been kept fixed. If we let $W_S$ vary along the structure we can achieve a maximum $\Delta\omega/2\pi$ of ~3.1GHz. But the strong variation of the Q values (80<Q<5000) will result in a strong change of the distribution of the dipole modes which will impact negatively on the wake. In this particular case, $W_S$ and $H_t$ are kept constant and single cell simulations indicate that the coupling stays almost the same from first to last cell.

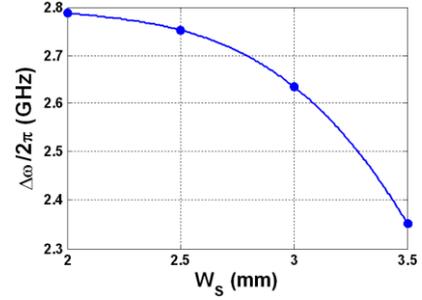

Figure 4: $\Delta\omega/2\pi$ as a function of $W_S$. In this particular case $W_S$ is kept constant throughout the all structure; for CLIC_DDS_A, $\Delta\omega/2\pi$ is ~2GHz (CLIC_DDS_C is ~2.3GHz [7]).

## WAKEFIELD SIMULATIONS

As a trade-off between frequency bandwidth and coupling, we have fixed the dimension of the first cell $W_S$=2.81mm and $H_S$=3mm ($H_t$=16.586mm throughout the full structure). We have considered a structure with 26 cells with a phase advance of $2\pi/3$ and we have spread the synchronous frequencies of the first dipole band in an erf distribution with a $\sigma$=814MHz ($\Delta\omega/2\pi$=2.685GHz), optimized for 6 RF cycles (0.5ns). Figure 5 shows the $\omega_{syn}/2\pi$ distribution as obtained from GdfidL simulations [14] in comparison with the uncoupled values obtained from symmetrical single cell simulations using HFSS [15].

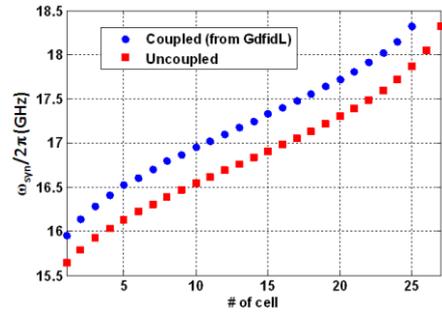

Figure 5: Synchronous frequency distribution in the structure.

It is worth noticing that the erf is built using 27 cells to take into account the presence of 27 irises and this gives a reduction of the bandwidth of ~230MHz from the actual structure with 26 cells. Further, coupling the cells results in a merging of two peaks which reduces the number of resonant peaks. The overall bandwidth is 2.36GHz, rather than the 2.47GHz specified in the uncoupled model. This bandwidth might be recovered by slightly increasing the

Q values which are shown along with the kick factors in Fig. 6.

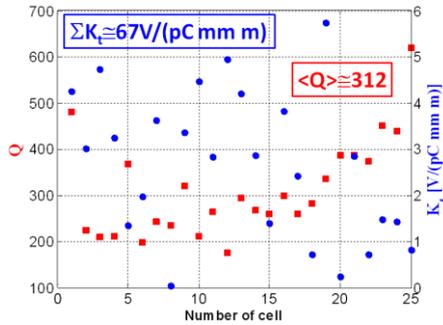

Figure 6: Q values and kick factors, $K_t$, obtained from GdfidL simulations of a single 26 cell structure.

Figure 7 illustrates the envelope of the transverse wake of the new design compared with CLIC_DDS_A. It should be noted that CLIC_DDS_A is a 24 cell structure, optimized to damping the first trailing bunch after 8 RF cycles (0.2m in the picture). Also the surface fields have not been optimized in the new design. However, as seen in the previous section, H-fields on the outer walls of the new design are intrinsically lower than in CLIC_DDS_A cells and we are reasonably confident that we can obtain a similar bandwidth and coupling once the surface fields have been optimized. This allows us to compare the two wakes.

The wakefield clearly represents an improved damping of the present design. However, further optimisation is necessary as the first trailing bunch is appreciable above the beam dynamics requirement of 6.7 V/pC/mm/m. In addition there are a few high Q modes which give rise to relatively flat wakefield profile. This is illustrated in Fig. 7 and the corresponding impedance spectrum is shown in Fig. 8.

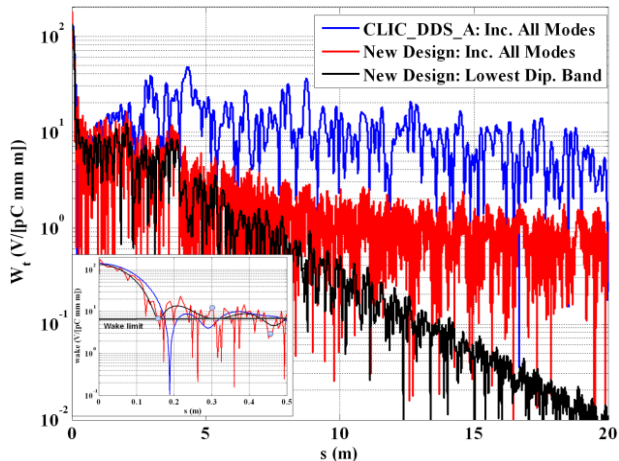

Figure 7: Transverse wake of a single structure, simulated with GdfidL using a quarter section. Shown inset is the wakefield of the first few trailing bunches (indicated by dots).

Interleaving the frequencies of successive structures will certainly improve the long-range damping properties. However, increasing the frequency bandwidth is the main method at our disposal to reduce the wakefield expressed by the first trailing bunch. This is facilitated by increasing the iris radii of the end cells. However, this will impact the surface electromagnetic fields adversely.

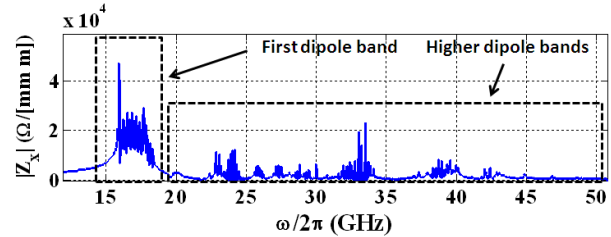

Figure 8: Impedance spectrum of the new 26 cell structure obtained from GdfidL simulations.

## FINAL REMARKS

The present wakefield design, although superior in damping properties to other ones, it does not meet the revised tighter beam dynamics constrains imposed for the CLIC main linacs.

A higher phase advance structure is currently being explored as a potential structure to reduce both the surface e.m. fields and the wakefields. This is based on a $5\pi/6$ phase advance per cell [7], [8].

## ACKNOWLEDGEMENTS

Research leading to these results has received funding from European commission under the FP7 research infrastructure grant no. 227579.